\documentclass[12pt]{aastex6}
\usepackage{epsfig}
\usepackage{url}
\usepackage{amsmath}
\usepackage{multirow}
\usepackage{mathptmx}
\usepackage{graphicx}

\begin{document}
\shorttitle{Internal origin of the GRB prompt GeV emission}
\shortauthors{Tang et al.}
\title{Evidence of internal dissipation origin for the high-energy prompt emission of GRB~170214A}
\author{Qing-Wen Tang\altaffilmark{1,2,3}, Xiang-Yu Wang\altaffilmark{4,5}, Ruo-Yu Liu$^{6}$}

\affiliation{$^{1}$Department of Physics, Nanchang University, Nanchang 330031, China }
\affiliation{$^{2}$Center for Cosmology and AstroParticle Physics (CCAPP),
The Ohio State University, Columbus, OH 43210, USA }
\affiliation{$^{3}$Guangxi Key Laboratory for Relativistic Astrophysics, Nanning 530004, China}
\affiliation{$^{4}$School of Astronomy and Space Science, Nanjing University, Nanjing 210093, China}
\affiliation{$^{5}$Key Laboratory of Modern Astronomy and Astrophysics (Nanjing University),
Ministry of Education, Nanjing 210093, China}
\affiliation{$^{6}$Max-Planck-Institut f\"ur Kernphysik, D-69117
Heidelberg, Germany}

\begin{abstract}
The origin of the prompt high-energy ($>100$MeV) emission of gamma-ray Bursts (GRBs),
detected by the Large Area Telescope (LAT) on board
the {\it Fermi} Gamma-ray Space Telescope, for which both an
external shock origin and internal dissipation origin have been suggested, is still under debate. In the internal
dissipation scenario, the high-energy emission is expected to exhibit significant temporal variability,
tracking the keV/MeV fast variable behavior. Here, we report a detailed analysis
of the {\it Fermi} data of GRB~170214A, which is sufficiently bright in the high energies to enable a quantitative analysis of the correlation between high-energy emission and keV/MeV emission with high statistics. Our result shows a
clear temporal correlation between high-energy and keV/MeV emission in the whole prompt emission phase as well as in two decomposed short time intervals.
Such a correlation behavior is also found in some other bright LAT GRBs, e.g., GRB 080916C, 090902B and 090926A. For these GRBs as well as GRB 090510, we also find the rapid temporal variability in the high-energy emission. We thus conclude that the prompt high-energy emission in these bright LAT GRBs should be due to an internal origin.
\end{abstract}
\keywords{gamma-ray burst: individual (GRB 170214A) ¨C radiation mechanisms: non-thermal}

\section{Introduction}
Gamma-ray bursts (GRBs) are the most intense astrophysical
explosions in the universe. The most popular model
for interpreting the highly variable keV-MeV emission, such as internal shocks, is
the internal dissipation model.
In recent years, {\it Fermi} Large Area
Telescope (LAT; 20~MeV to more than 300~GeV) has detected prompt and
long-lived high-energy ($>100$~MeV) gamma-ray emissions
from a large number of GRBs, such as GRB~080916C, GRB~090510,
GRB~090902B, GRB~090926A and GRB~130427A~\citep{2009Sci...323.1688A,2010ApJ...716.1178A,2009ApJ...706L.138A,2011ApJ...729..114A,2014Sci...343...42A}.
The long-lived high-energy emissions are believed to be produced by the external shocks~\citep{2009MNRAS.400L..75K,2010MNRAS.409..226K,2010ApJ...720.1008C,2010ApJ...709L.146D,2009ApJ...706L..33G,2010A&A...510L...7G,2010MNRAS.403..926G,2010ApJ...712.1232W,2010ApJ...724L.109R},
via synchrotron emission and/or inverse-Compton processes.
However, the origin of the high-energy photons during the prompt phase
is still uncertain. It has been suggested that the prompt high-energy emission
also arises from the external shocks, via synchrotron radiation (e.g., \citet{2009MNRAS.400L..75K})
or scattering prompt MeV photons by the accelerated electrons there~\citep{2014ApJ...788...36B}.
Such external origin models predict a smooth light curve of high-energy emission.
On the other hand, there have been indications of the internal origin of
the prompt high-energy emissions for some GRBs, such as GRB~090926A and GRB~090902B,
as prompt high-energy emissions show a variable structure correlating with the keV-MeV
emission~\citep{2011ApJ...729..114A,2011ApJ...730..141Z}. If such a temporal behavior of
high-energy emission is real, it would favor the internal origin scenario.

Recently, {\it Fermi}-LAT observed a bright GRB~170214A, with
more than one hundred of $>100$~MeV photons within the first
200 seconds, which makes it a good case for studying the
temporal correlation in a statistical way. In this work, we present a quantitative analysis of
the prompt variable keV-MeV and high-energy emissions of GRB~170214A,
and compare it with other bright LAT GRBs.

\section{Data analysis}
\label{data}
\subsection{Properties of GRB 170214A}
{\it Fermi} Gamma-Ray Burst Monitor (GBM, energy coverage of 8~keV-40~MeV) is triggered
by GRB~170214A at $T_0=$15:34:26.92 UT on 14 February 2017 ($T_0$, the GBM trigger time).
The GBM light curve shows multiple overlapping peaks with a $T_{90}$
duration of about $123$ seconds~\citep{2017GCN..20675...1M}. Simultaneously, {\it Fermi}-LAT
detected high-energy emission from GRB 170214A, at a location of
R.A.$=256.33$, decl.$=-1.88$ (J2000)~\citep{2017GCN..20687...1M}, which is consistent with
that detected by {\it Fermi}-GBM. More than $160$ photons above $100$~MeV,
with $13$ of them above $1$~GeV, are observed within $1000$ seconds~\citep{2017GCN..20687...1M},
which makes it a good case to perform time-resolved analysis of high-energy emission.

The Konus-Wind detected the multi-peak lightcurve with a $T_{90}$ duration
of about $150$ seconds~\citep{2017GCN..20678...1F}. {\it Swift}-XRT detected an afterglow emission
close to the LAT position~\citep{2017GCN..20691...1B,2017GCN..20679...1B}. Follow-up observations in
the optical and/or NIR band are performed by RATIR (the Reionization and Transients
Infrared Camera), NOT (the Nordic Optical Telescope), GROND and
Mondy (the AZT-33IK telescope in Sayan observatory)~\citep{grb-ratir,grb-ratir2,2017GCN..20683...1M,grb-grond,2017GCN..20684...1S,2017GCN..20687...1M}.
The ESO Very Large Telescope detected a faint optical afterglow and claimed
a redshift of $z=2.53$~\citep{2017GCN..20686...1K}.

\subsection{LAT data analysis}
Within $12$ degrees of the reported LAT position, R.A.$=256.33$, decl.$=-1.88$ (J2000)~\citep{2017GCN..20687...1M}, the Pass 8 transient
events are used in the energy range of
$100$~MeV to $10$~GeV. These data are analyzed using
the {\it Fermi} ScienceTools package (v10r0p5)
available from the {\em Fermi} Science Support Center(FSSC) \footnote{\url{https://fermi.gsfc.nasa.gov/ssc/}}.
Events with zenith angles $>$100$^\circ$ are excluded to reduce the contribution
of Earth-limb gamma rays. Instrument response function ``P8R2$\_$TRANSIENT020$\_$V6'' is used.
GRB 170214A is modeled as a point source with the corresponding position
and the photon spectrum is assumed to be a power law, i.e.,
$dN/dE=N_0 (E/{\rm 100MeV})^{-\Gamma_{\rm LAT}}$, with the
normalization factor ($N_0$) and photon index ($\Gamma_{\rm LAT}$) as free parameters.
A background model comprises the galactic interstellar
emission model (``gll$\_$iem$\_$v06.fits'') and extragalactic isotropic
spectral template (``iso$\_$P8R2$\_$TRANSIENT020$\_$V6$\_$v06.txt'').
For these diffuse components in the model, we calculate the response files
by the {\it gtdiffrsp} tool. The livetime cube and exposure maps are
generated by the {\it gtltcube} and {\it gtexpmap} tool. We run the {\it gtlike}
tool to derive the best fit.

We first perform a blind search in three good time intervals,
i.e., $0-900$, $2500-7000$ and $8500-13000$ second after
the GBM trigger. The strong emission exists in the first $900$ seconds,
after which no significant emission is found.

Second, $10$ and $100$ seconds are employed as the resolved time bin
before and after $T_0+200$ seconds respectively, when performing the
time-resolved analysis of the first $900$ seconds data. The nearby time bin
is combined if the error of the energy flux in one time bin is larger than
then central value. For the intensive emission period, i.e., 52-70~s,
we divide this time interval into 7 bins. Before 52~s,
the LAT show a marginal significant emission, which is treated as a single time bin.
The likelihood results, i.e., the photon flux ($F_{\rm L}$) and
energy flux ($f_{\rm L}$) are present in Tab.~\ref{tab1},
where the total photon number within $12$ degrees ($N_{\rm ROI}$), the predicted photon
number ($N_{\rm P}$) and the the test-statistic value (TS, the square root of TS approximately equals to the detection significance~\citep{1996ApJ...461..396M}), are also given.

\begin{table*}
\centering
\caption{{\em Fermi}-LAT likelihood results for GRB 170214A. \label{tab1}}
\begin{tabular}{ccccccc}
\hline
$T_1$	-	$T_2$\tablenotemark{\it a}	&	TS\tablenotemark{\it b}	&	$N_{\rm ROI}$\tablenotemark{\it c}	
&	$N_{\rm P}$\tablenotemark{\it d}	&	$\Gamma_{\rm LAT}$\tablenotemark{\it e}			&	 $F_{\rm L} (0.1-10 {\rm GeV})$\tablenotemark{\it e}  &	$f_{\rm L} (0.1-10 {\rm GeV})$\tablenotemark{\it e}			\\	
	s		&		&		&		&				&	${\rm 10^{-5} ph \ cm^{-2}\ s^{-1}}$			 
&	${\rm 10^{-8} erg \ cm^{-2}\ s^{-1}}$			 \\	\hline
0	-	52	&	15 	&	16	&	7.2 	&	4.98 	$\pm$	2.62 	&	3.01 	$\pm$	1.34 	
&	 0.64 	$\pm$	0.34 	\\	
52	-	62	&	11 	&	4	&	4.0 	&	6.96 	$\pm$	2.76 	&	8.92 	$\pm$	4.44 	
&	 1.71 	$\pm$	0.86 		\\	
62	-	63	&	104 	&	15	&	14.4 	&	3.04 	$\pm$	0.54 	&	285.06 	$\pm$	89.54 	 
&	88.94 	$\pm$	29.37 	\\	
63	-	64	&	308 	&	16	&	16.0 	&	4.05 	$\pm$	0.70 	&	333.88 	$\pm$	182.45 	 
&	79.51 	$\pm$	51.58 		\\	
64	-	66	&	38 	&	10	&	8.2 	&	3.73 	$\pm$	0.91 	&	84.86 	$\pm$	35.81 	
&	 21.44 	$\pm$	9.35 		\\	
66	-	67	&	20 	&	5	&	5.0 	&	3.93 	$\pm$	1.21 	&	103.80 	$\pm$	46.62 	
&	 25.21 	$\pm$	12.01 		\\	
67	-	69	&	57 	&	6	&	5.9 	&	2.78 	$\pm$	0.73 	&	57.75 	$\pm$	30.29 	
&	 20.53 	$\pm$	11.44 		\\	
69	-	70	&	19 	&	5	&	3.7 	&	3.19 	$\pm$	1.06 	&	74.45 	$\pm$	40.72 	
&	 21.82 	$\pm$	13.74 		\\	
70	-	80	&	58 	&	11	&	10.0 	&	3.50 	$\pm$	0.79 	&	20.37 	$\pm$	7.11 	
&	 5.43 	$\pm$	2.26 		\\	
80	-	90	&	196 	&	11	&	11.0 	&	2.50 	$\pm$	0.42 	&	20.65 	$\pm$	6.29 	 
&	8.95 	$\pm$	3.82 		\\	
90	-	100	&	74 	&	10	&	9.3 	&	3.31 	$\pm$	0.70 	&	18.65 	$\pm$	6.65 	
&	 5.26 	$\pm$	2.07 		\\	
100	-	110	&	58 	&	6	&	5.2 	&	1.54 	$\pm$	0.29 	&	7.89 	$\pm$	3.79 	
&	 11.89 	$\pm$	6.55 		\\	
110	-	120	&	76 	&	11	&	9.5 	&	2.37 	$\pm$	0.43 	&	16.90 	$\pm$	6.02 	
&	 8.25 	$\pm$	3.96 		\\	
120	-	130	&	92 	&	12	&	11.1 	&	2.01 	$\pm$	0.30 	&	18.66 	$\pm$	6.11 	
&	 13.81 	$\pm$	6.26 		\\	
130	-	140	&	236 	&	13	&	12.9 	&	1.92 	$\pm$	0.25 	&	21.39 	$\pm$	6.04 	 &	17.85 	$\pm$	7.34 		\\	
140	-	150	&	97 	&	16	&	13.0 	&	2.20 	$\pm$	0.32 	&	21.91 	$\pm$	6.38 	
&	 12.75 	$\pm$	5.36 		\\	
150	-	160	&	117 	&	9	&	8.9 	&	2.38 	$\pm$	0.44 	&	15.35 	$\pm$	5.59 	 
&	7.38 	$\pm$	3.63 		\\	
160	-	170	&	108 	&	7	&	7.0 	&	2.35 	$\pm$	0.47 	&	11.82 	$\pm$	4.51 	 
&	5.88 	$\pm$	3.28 		\\	
170	-	180	&	73 	&	6	&	6.0 	&	2.10 	$\pm$	0.43 	&	9.37 	$\pm$	3.88 	
&	 6.14 	$\pm$	3.79 		\\	
180	-	190	&	147 	&	8	&	8.0 	&	2.27 	$\pm$	0.42 	&	12.86 	$\pm$	4.60 	 
&	6.91 	$\pm$	3.65 		\\	
190	-	200	&	54 	&	6	&	6.0 	&	2.60 	$\pm$	0.60 	&	9.89 	$\pm$	4.07 	
&	 3.95 	$\pm$	2.22 		\\	
200	-	300	&	142 	&	43	&	39.7 	&	2.67 	$\pm$	0.27 	&	6.43 	$\pm$	1.17 	 
&	2.45 	$\pm$	0.56 		\\	
300	-	500	&	140 	&	55	&	33.5 	&	2.24 	$\pm$	0.21 	&	2.58 	$\pm$	0.49 	 
&	1.43 	$\pm$	0.38 		\\	
500	-	700	&	65 	&	38	&	20.8 	&	2.50 	$\pm$	0.32 	&	1.64 	$\pm$	0.41 	
&	 0.71 	$\pm$	0.23 		\\	
700	-	900	&	22 	&	19	&	7.7 	&	2.21 	$\pm$	0.43 	&	0.59 	$\pm$	0.25 	
&	 0.34 	$\pm$	0.20 		\\	
\hline
190	-	900	&	380 	&	161	&	106.2 	&	2.44 	$\pm$	0.14 	&	2.37 	$\pm$	0.54 	 
&	1.07 	$\pm$	0.36 		\\
\hline
\end{tabular}
\tablenotetext{\it a}{The start analysis time ($T_1$) and the end
analysis time ($T_2$) in unit of seconds.}
\tablenotetext{\it b}{TS is the test-statistic value, which is roughly equal to $\sigma^2$,
where $\sigma$ is the significance of GRB detection.}
\tablenotetext{\it c}{The observed LAT counts number
within the region of interest (ROI), i.e., 12 degrees of GRB center position.}
\tablenotetext{\it d}{The predicted LAT counts number from GRB 170214A.}
\tablenotetext{\it e}{Photon index ($\Gamma_{\rm LAT}$), photon flux ($F_{\rm L}$) and energy flux ($f_{\rm L}$) of GRB 170214A.}
\end{table*}

\subsection{GBM data analysis}
Given the recommendation for selecting the detectors with
high counts rate above background, the Time Tagged Event (TTE) data from two NaI detectors (n0, n1)
and one BGO detector (b0) are taken from FSSC and analyzed
with the software package RMFIT version 4.3pr2 \footnote{\url{https://fermi.gsfc.nasa.gov/ssc/data/analysis/rmfit/}}.
We select the energy range of $8-1000$~keV for two NaI detectors and
of $200$~keV--$10$~MeV for a BGO detector. A first order polynomial is applied for
each detector to fit the background with flat counts rate regions pre- and post-burst.

We select the time bins same as that used in LAT analysis in the time interval
of $52-160$~s, after which there is no significant emission in GBM band.
Before $52$~s, we perform a small time bins, i.e., 2 seconds per time bin,
since it is a fast variable (hereafter FV) component in GBM energy bands\footnote{We are actually unable to recognize every single pulse with very short time variability in the light curves of energy flux (e.g., $\sim 100$~ms). However, we can still search temporal correlation in a longer timescale. The light curves of both the LAT emission and GBM emission show structures with a fast-rising followed by a fast-decaying in the time interval 52-80~s and 90-160~s respectively (see next section), so we define these structures as fast variable (FV) components to search the correlation in two energy bands.}. The Band function is employed
as the photon spectrum model in each time bin, which is described by~\citet{1993ApJ...413..281B}:
\begin{displaymath}
N(E)= A\left\{ \begin{array}{ll}
(E/100\,\mathrm{keV})^\alpha e^{(-E(2+\alpha)/E_p)} & \textrm{if $E<E_b$}\\
{[(\alpha - \beta) E_p/(100\,\mathrm{keV}(2+\alpha))]}^{(\alpha-\beta)}
e^{\beta-\alpha} (-E/100\,\mathrm{keV})^\beta & \textrm{if $E \geq
E_b $}
\end{array} \right.
\end{displaymath}
where $A$ is the normalization, $E_b=(\alpha - \beta) E_p/(2+\alpha)$, $\alpha$ is the
photon index at low energy, $\beta$ is the photon index at high
energy and $E_p$ is the peak energy in the $E^2 N(E)$
representation.
The energy fluxes ($f_{\rm G}$)
are obtained between $10$~keV and $10$~MeV, as shown in Tab.~\ref{tab2}.

For GRB 170214A, we build the light curves in two narrow energy bands, i.e., 8-200~keV and 200~keV-1~MeV. We employ a single power-law function (PL), i.e., $N(E)=N_0 (E/{\rm 100keV})^{-\alpha_{\rm PL}}$ ($\alpha_{\rm PL}$, the PL decay index),  to model the photon spectrum in the former energy band, because the derived peak energy in the whole GBM energy range (8~keV-10~MeV) is always larger than $\sim 200$~keV. A Band function is used to model the photon spectrum for the latter energy band. The GBM fluxes in these two energy bands of GRB 170214A are present in Tab.~\ref{tab2}.

\begin{table*}
\centering
\caption{{\it Fermi}-GBM results for GRB 170214A. \label{tab2}}
\begin{tabular}{cccc}
    \hline
$T_1$	-	$T_2$\tablenotemark{\it a}  &   $f_{\rm G} (10 {\rm keV}-10 {\rm MeV})$\tablenotemark{\it b}  &   $f_{\rm G} (200 {\rm keV}-1 {\rm MeV})$\tablenotemark{\it b}    &   $f_{\rm G} (8 {\rm keV}-200 {\rm MeV})$\tablenotemark{\it b}  \\
 s      &   ${\rm 10^{-7} erg \ cm^{-2}\ s^{-1}}$  &${\rm 10^{-7} erg \ cm^{-2}\ s^{-1}}$  &${\rm 10^{-7} erg \ cm^{-2}\ s^{-1}}$  \\
    \hline
    0	-	2	&	8.23	$\pm$	1.60	&	3.78	$\pm$	0.61	&	1.73	$\pm$	0.13	 \\
2	-	4	&	14.70	$\pm$	3.79	&	3.69	$\pm$	0.84	&	2.27	$\pm$	0.14	\\
4	-	6	&	21.40	$\pm$	4.43	&	3.89	$\pm$	0.12	&	2.99	$\pm$	0.15	\\
6	-	8	&	18.80	$\pm$	4.80	&	7.92	$\pm$	0.77	&	3.31	$\pm$	0.15	\\
8	-	10	&	27.60	$\pm$	5.01	&	10.20	$\pm$	0.75	&	4.10	$\pm$	0.15	\\
10	-	12	&	23.80	$\pm$	4.99	&	9.41	$\pm$	0.73	&	4.18	$\pm$	0.15	\\
12	-	14	&	29.40	$\pm$	4.82	&	12.60	$\pm$	0.82	&	5.77	$\pm$	0.17	\\
14	-	16	&	27.40	$\pm$	1.81	&	12.10	$\pm$	0.72	&	5.67	$\pm$	0.16	\\
16	-	18	&	35.40	$\pm$	5.14	&	16.00	$\pm$	2.68	&	6.95	$\pm$	0.17	\\
18	-	20	&	37.00	$\pm$	1.76	&	16.50	$\pm$	0.82	&	6.81	$\pm$	0.18	\\
20	-	22	&	41.40	$\pm$	5.00	&	18.40	$\pm$	0.84	&	8.06	$\pm$	0.18	\\
22	-	24	&	27.10	$\pm$	4.62	&	12.40	$\pm$	0.77	&	7.02	$\pm$	0.17	\\
24	-	26	&	16.40	$\pm$	0.92	&	7.25	$\pm$	0.67	&	5.32	$\pm$	0.15	\\
26	-	28	&	16.30	$\pm$	1.44	&	$<$10.3                	&	4.36	$\pm$	0.15	\\
28	-	30	&	20.50	$\pm$	4.00	&	10.10	$\pm$	0.71	&	6.69	$\pm$	0.16	\\
30	-	32	&	24.30	$\pm$	4.26	&	8.00	$\pm$	0.70	&	5.48	$\pm$	0.16	\\
32	-	34	&	20.80	$\pm$	4.21	&	9.37	$\pm$	0.70	&	6.24	$\pm$	0.16	\\
34	-	36	&	38.00	$\pm$	2.01	&	16.30	$\pm$	0.85	&	6.75	$\pm$	0.17	\\
36	-	38	&	39.00	$\pm$	4.92	&	20.00	$\pm$	0.89	&	8.85	$\pm$	0.18	\\
38	-	40	&	39.60	$\pm$	1.87	&	16.80	$\pm$	1.37	&	8.35	$\pm$	0.17	\\
40	-	42	&	39.80	$\pm$	1.88	&	17.80	$\pm$	0.97	&	7.08	$\pm$	0.18	\\
42	-	44	&	55.00	$\pm$	5.00	&	25.00	$\pm$	0.93	&	9.89	$\pm$	0.19	\\
44	-	46	&	68.70	$\pm$	5.41	&	26.50	$\pm$	2.87	&	9.58	$\pm$	0.20	\\
46	-	48	&	54.00	$\pm$	5.15	&	19.20	$\pm$	1.02	&	7.73	$\pm$	0.19	\\
48	-	50	&	47.50	$\pm$	2.04	&	20.30	$\pm$	1.13	&	8.28	$\pm$	0.18	\\
50	-	52	&	68.40	$\pm$	4.10	&	23.80	$\pm$	0.90	&	9.72	$\pm$	0.20	\\
52	-	62	&	39.00	$\pm$	1.89	&	15.60	$\pm$	0.34	&	7.66	$\pm$	0.08	\\
62	-	63	&	133.00	$\pm$	6.13	&	33.70	$\pm$	1.51	&	12.10	$\pm$	0.29	\\
63	-	64	&	98.10	$\pm$	5.99	&	27.90	$\pm$	1.70	&	9.93	$\pm$	0.27	\\
64	-	66	&	59.80	$\pm$	3.90	&	20.60	$\pm$	0.83	&	9.55	$\pm$	0.19	\\
66	-	67	&	56.20	$\pm$	5.78	&	19.10	$\pm$	1.25	&	8.86	$\pm$	0.25	\\
67	-	69	&	47.00	$\pm$	3.84	&	14.90	$\pm$	0.81	&	7.91	$\pm$	0.18	\\
69	-	70	&	29.40	$\pm$	4.42	&	$<$16.7	                &	5.01	$\pm$	0.21	\\
70	-	80	&	28.90	$\pm$	1.82	&	12.70	$\pm$	0.60	&	7.19	$\pm$	0.08	\\
80	-	90	&	7.02	$\pm$	1.58	&	$<$5.1	                &	2.16	$\pm$	0.06	\\
90	-	100	&	2.74	$\pm$	1.39	&	0.44	$\pm$	0.25	&	1.24	$\pm$	0.05	\\
100	-	110	&	6.97	$\pm$	1.25	&	1.83	$\pm$	0.42	&	2.25	$\pm$	0.06	\\
110	-	120	&	4.44	$\pm$	0.61	&	$<$2.2	                &	0.90	$\pm$	0.05	\\
120	-	130	&	9.33	$\pm$	1.36	&	1.71	$\pm$	0.30	&	1.86	$\pm$	0.06	\\
130	-	140	&	22.80	$\pm$	1.80	&	7.61	$\pm$	0.32	&	5.65	$\pm$	0.07	\\
140	-	150	&	6.08	$\pm$	1.11	&		$<$4.2	            &	1.97	$\pm$	0.06	\\
150	-	160	&	2.16	$\pm$	1.06	&	    $<$1.2	            &	0.24	$\pm$	0.06	\\
    \hline
\end{tabular}
\tablenotetext{\it a}{The start analysis time ($T_1$) and the end
analysis time ($T_2$) in unit of seconds.}
\tablenotetext{\it b}{GBM energy flux in corresponding energy range.}
\end{table*}

\section{Temporal and spectral analysis of GRB 170214A}

\subsection{Light curves}

\begin{figure*}
\centering
\includegraphics[height=6cm]{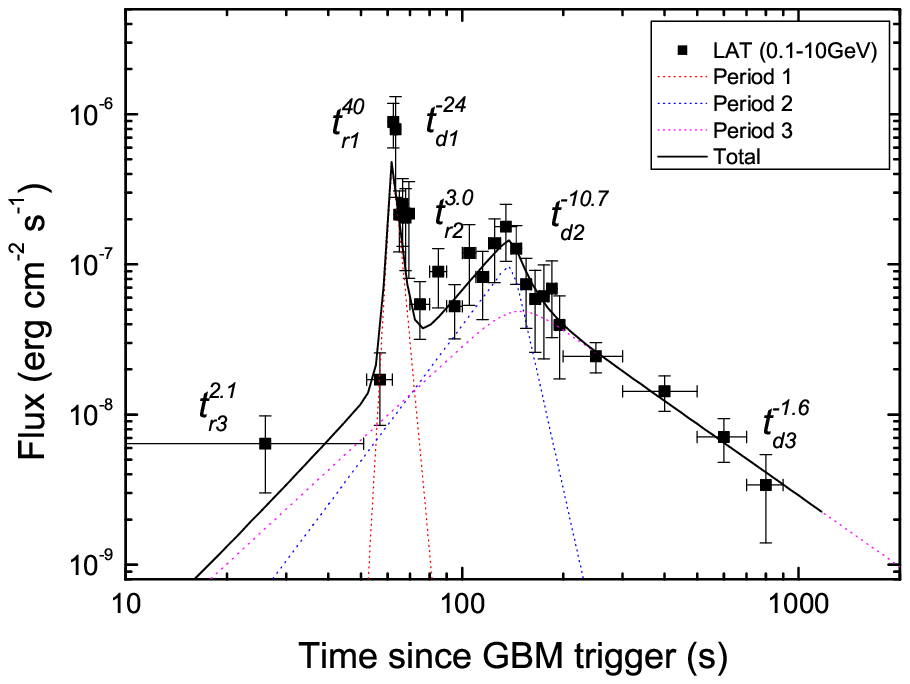}
\includegraphics[height=6cm]{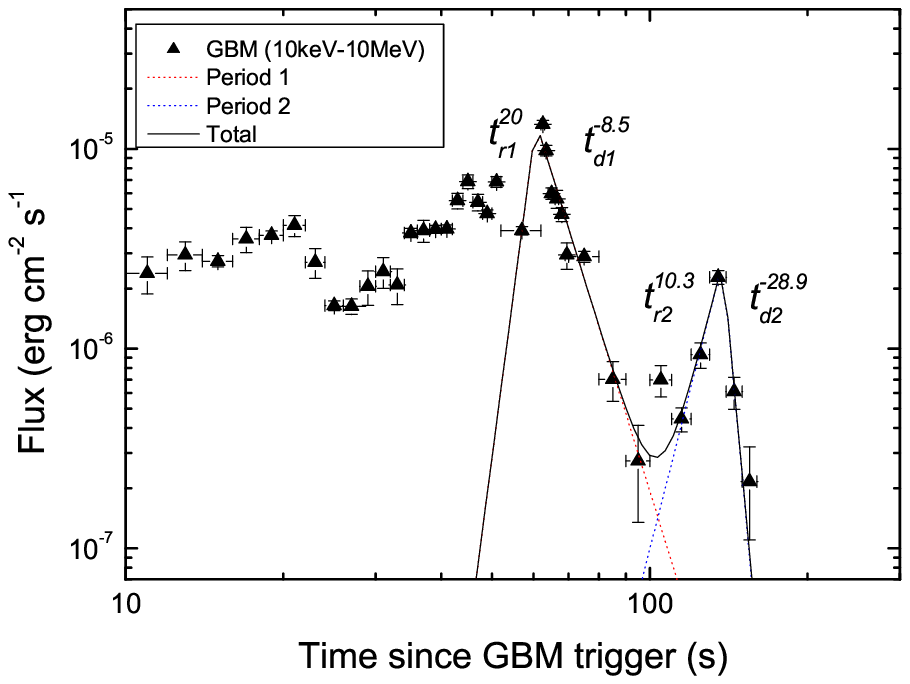}
\caption{The LAT and GBM light curves of  GRB 170214A
with the best fits to these data. The rising and decay indices ($\alpha_r$ and $\alpha_d$) for each component are labeled in corresponding position. \label{fig1}}
\end{figure*}

The light curves of GRB 170214A in LAT (0.1-10~GeV) and GBM (10~keV-10~MeV) bands are plotted in Fig.~\ref{fig1}, which can be described with four phases.
\begin{enumerate}
\item {\it 0-52~s}: The LAT emission is too weak to be subdivided in this
    period, while the GBM emission show a few pulse structures.

\item {\it 52-80~s} (Period 1): One fast variable (FV) component is found at
    both the LAT and GBM bands with fast rising and fast decaying behaviors. We fit it with an empirical smooth broken power law function (SBPL):
    \begin{equation}
    f(t)=f_0 [(\frac{t}{t_p})^{-\alpha_r s}+(\frac{t}{t_p})^{-\alpha_d s}]^{-1/ s}
    \end{equation}
    where $f_0$ is the normalisation in unit
    of ${\rm erg\ cm^{-2}\ s^{-1}}$, $t_p$ is the peak time
    of the FV component, and the temporal indices
    of the rising part and the decaying part
    are $\alpha_r$ and $\alpha_d$ respectively. Here $s$ determines the smoothness of the peak, which is fixed at $10$ for the fast variable components, following the suggestions in~\citet{2008ApJ...675..528L}.
    For both the GBM and LAT bands, the value of
    $\alpha_r$ can not be constrained given only two flux points in the rising part, and hence we fix their values based on the connection of the two data points
    in the two energy bands respectively, i.e., $\alpha_r=40$ in LAT band and $\alpha_r=20$ in GBM band. As shown in Tab.~\ref{tab3}, both light curves decay steeply, i.e., $\alpha_d=-24\pm14.5$ in LAT band and $\alpha_d=-8.5\pm0.4$ in GBM band. Although the error bar of $\alpha_d$ is quite large in the LAT band, the result clearly shows a quick flux drop after the peak. Such strong variabilities imply that they are mostly likely to be related with central engine of GRB 170214A. The peak times $t_p$ in both energy bands are consistent with each other with uncertainties, which are around 61.8 seconds after GBM trigger.

\item {\it 90-160~s} (Period 2): One FV component appears in either the light curve both in the GBM or LAT band.
    By fitting them with a SBPL function, $\alpha_r$ is found to be $3.0\pm1.1$ for
    the LAT emission and $10.3\pm1.0$ for the GBM emission, which are a bit too rapid
    for the external reverse shock model. This is also proved
    by the large decay indices ($\alpha_d$) in both energy bands, which
    are $-10.7\pm7.4$ and $-28.9\pm17.8$ for LAT and GBM emission respectively, although the resultant error bars are quite large.
    The peak times in both energy bands are consistently around 139 seconds after GBM trigger.

\item {\it 160-900~s} (Extended phase): No GBM emission is detected in this phase,
    which implies GRB 170214A enters the so called afterglow phase.
    The LAT light curve shows a power law decay with decay index ($\alpha_{\rm LAT}$) of $-1.6\pm0.2$.
    The LAT photon index ($\Gamma_{\rm LAT}$) in the time
    interval (190-900~s) is found to be $-2.4\pm0.1$,
    translating to a spectral index $\beta_{\rm LAT} = \Gamma_{\rm LAT} + 1 = -1.4\pm0.1$.
    In the external shock model, we have the synchrotron flux at high energy $f_{LAT} \propto \nu^\beta t^\alpha$.
    Considering that the LAT energy band ($>100 {\rm MeV}$) is usually above
    the external synchrotron cooling energy ($h \nu_c$, $h$ is the plank constant),
    we can derive the injection electron spectrum power
    index $p$ to be $\sim2.8$ from $f_{LAT} \propto \nu^{-p/2}$. This predicts
    a power-law index of about $-1.6$ for the light
    curve in the external shock model, i.e., $f_{LAT} \propto t^{(2-3p)/4}$~\citep{1998ApJ...497L..17S}, which
    is consistent with the observed one, that is $-1.6\pm0.2$.
    Thus, we conclude that the late LAT emission can be well
    explained by the external shock model.
\end{enumerate}

Second, we perform a global fit to the LAT light curve in the whole detection interval,
i.e., 0-900 seconds. Based on above analysis, we decompose
three components from the LAT light curve, i.e., the Period 1, Period 2 and an underlying component, Period 3, with each of them modeled by a SPBL. The results are present in Tab.~\ref{tab3} and plotted in Fig.~\ref{fig1}.
As for the Period 1 and Period 2, the results are similar to that discussed above.
For the Period 3 (with sharpness $s$ of $3$), the peak time is around $145$~s, which can be explained as the dynamic
deceleration time of the ejecta. The rising temporal index is around $2.1$, which is consistent
with the expected index ($\sim 2.0$) in the external shock model~\citep{2009MNRAS.400L..75K,2010MNRAS.403..926G}.
Apparently, the flux of the Period 3 is comparable with that of
the Period 2 at $\sim$ 160.8 seconds after GBM trigger,
after which the afterglow emission takes over.
This time is also consistent with the end time of GBM emission.

\begin{table*}
\centering
\caption{Temporal behaviors of GRB 170214A in LAT and GBM band. \label{tab3}}
\begin{tabular}{cc|cccc|cccc}
\hline
&   &   &LAT    &   &   &   &GBM    &   &   \\
\hline
$T_1$	-	$T_2$	&	Period	&	$\alpha_r$\tablenotemark{\it a}			&	 $\alpha_d$\tablenotemark{\it a}			&	$t_p$\tablenotemark{\it a}	& $s$\tablenotemark{\it a}
&	 $\alpha_r$\tablenotemark{\it a}			&	$\alpha_d$\tablenotemark{\it a}			&	 $t_p$\tablenotemark{\it a}		&$s$\tablenotemark{\it a}	\\
s			&		&	&	&	s		&	&	&	&	s	&		\\    \hline
52	-	80	&	1	&	40	(fixed)		&	-24	$\pm$	14.5  	&	62.3	$\pm$	1.1	&10   &	20	 (fixed) &	-8.5	$\pm$	0.4	&	60.7	$\pm$	0.4 	&10\\
90	-	160	&	2	&	3.0 	$\pm$	1.1	&	-10.7	$\pm$	7.4	&	140.5	$\pm$	14.9	&10
&	10.3	$\pm$	1.0 	
&	-28.9	$\pm$	17.8	&	137.8	$\pm$	3.4	   &10\\
0	-	900	&	3	&	2.1	$\pm$	0.4	&	-1.6	$\pm$	0.2	&	145	$\pm$	21.9	 &3  &	-
&	-			&	-		&	\\
\hline
\end{tabular}
\tablenotetext{\it a}{The parameters of smoothly broken power-law function (SBPL), $\alpha_r$ is the rising index before the peak time of $t_p$, after which the temporal decay index is $\alpha_d$, $s$ is the smoothness of the break.}
\end{table*}

\subsection{Spectral analysis}
\begin{figure*}
   \centering
         \includegraphics[height=5cm]{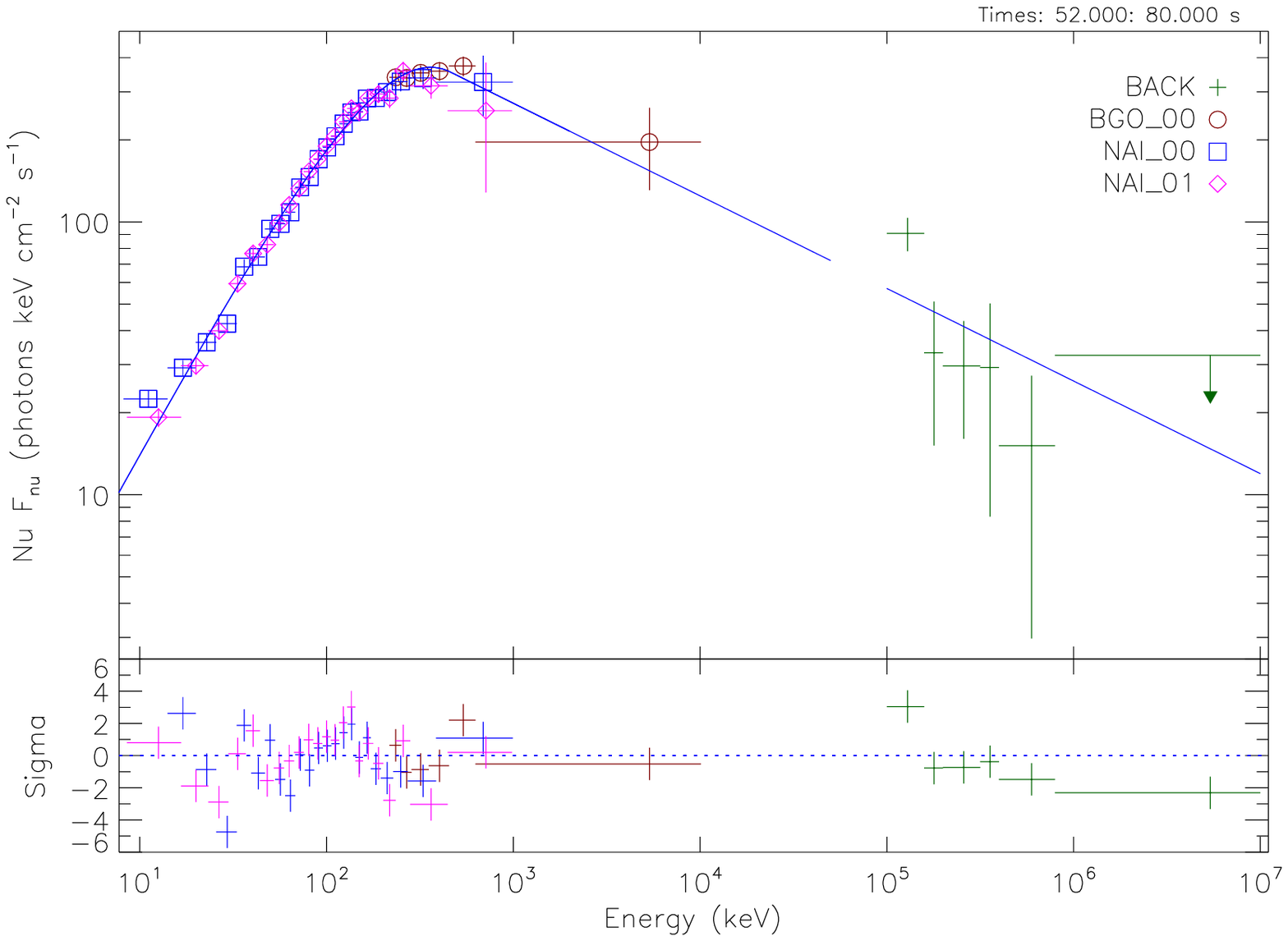}
         \includegraphics[height=5cm]{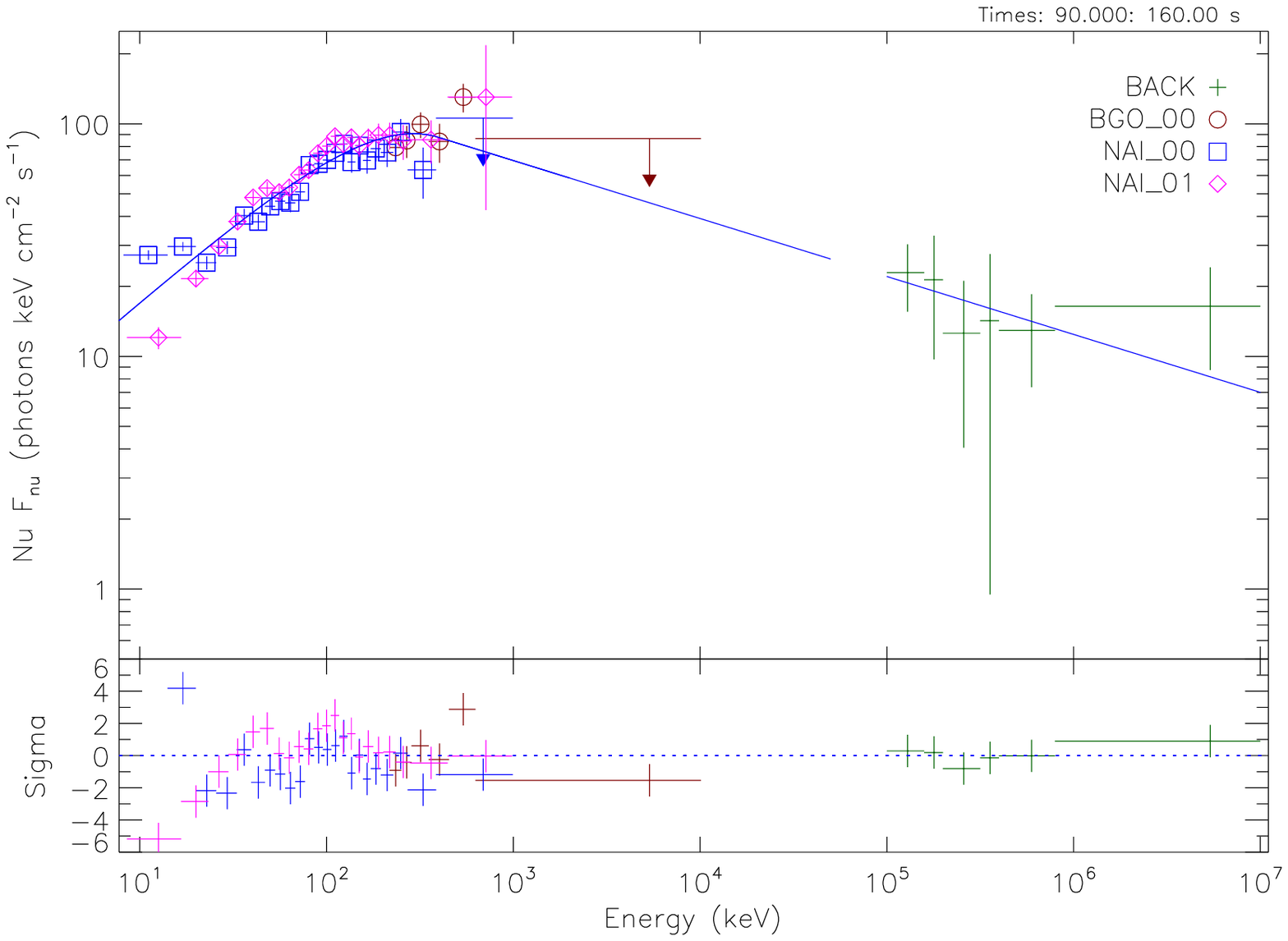}
      \caption{The GBM+LAT joint energy spectrum fits of GRB 170214A in two periods, which are modeled as the ``Band'' function~\citep{1993ApJ...413..281B}. For the former, $\alpha=-0.74\pm0.02$, $\beta=-2.34\pm0.06$ and $E_p=361\pm12$ keV, and for the latter, $\alpha=-1.30\pm0.03$, $\beta=-2.25\pm0.11$ and $E_p=292\pm37$ keV. \label{fig2}}
   \end{figure*}

In the Period 1 and Period 2, we perform a joint spectral fit employing the GBM and LAT data between 8~keV and 10~GeV. The Castor Statistic (CSTAT) is used in the spectral fit, as in other bright LAT GRBs~\citep{2009ApJ...706L.138A,2010ApJ...716.1178A,2011ApJ...729..114A}. For the same degrees of freedom (DOF), the smaller the CSTAT value is,  the better the photon model is for the data.
The results are present in Tab.~\ref{tab4}, where the photon index $\Gamma_{\rm LAT}$ that derived from the LAT data only is also presented to compare with results of the GBM+LAT joint fit. The results are also shown in Fig.~\ref{fig2}.

For the intensive period of 52-80~s,
the Band parameters, $\alpha$, $\beta$ and $E_p$, are found to be $-0.74\pm0.02$,
$-2.34\pm0.06$ and $361\pm12$ keV respectively.
The CSTAT value is 600 with DOF of 348.
$\Gamma_{\rm LAT}$ from the LAT data only is much softer than $\beta$ from the GBM+LAT joint fit in this period, i.e., $-\Gamma_{\rm LAT}$ of $-3.49\pm0.28$ comparing with $\beta$ of $-2.34\pm0.06$.
Thus, we test a BandCut model, which is
described as the Band with a high energy cutoff, i.e., $e^{-E/E_c}$, where $E_c$ is the exponential cutoff energy.
With one more parameter, the BandCut will be regarded as a more preferred model
than a single Band if $\Delta$(CSTAT) is larger than
28~\citep{2013ApJS..209...11A,2015ApJ...806..194T}. However, only $\Delta$(CSTAT)$\sim$25
is found for the BandCut model with a cutoff energy of $224\pm58$~MeV.
Therefore, we consider they are the equally good models in this period.
Alternatively, when a power law function
is added to the Band model, i.e., Band+PL, the fit becomes even worse than a single Band with a
larger CSTAT value.

For the second period of 90-160~s,
$\alpha$, $\beta$ and $E_p$ in the Band model are -1.30$\pm$0.03,
$-2.25\pm0.11$ and $292\pm37$ keV respectively, with the CSTAT/DOF of
714/348. The photon index of the LAT data is 2.12$\pm$0.13,
which is consistent with the high-energy photon index $\beta$ of the GBM+LAT joint fit.
The spectrum fit in this period cannot be improved significantly
by either employing another fitting function or adding an additional
component to the Band function, i.e., with a larger CSTAT value.

The result implies the prompt GBM emission and the prompt LAT emission of GRB 170214A may come from the same region, and  disfavors the existence of other spectral components in these two periods.

\begin{table*}
\centering
\caption{Spectral analysis results of GRB 170214A in two periods. \label{tab4}}
\begin{tabular}{c|ccclcc|cccc}
    \hline
       &   &GBM+LAT  &  &   &     &    &LAT    &   &   \\
       &   &8keV-10~GeV  &  &   &     &    &0.1-10~GeV    &   &   \\
    \hline
$T_1$	-	$T_2$		&Model\tablenotemark{\it a}    &	$\alpha$\tablenotemark{\it a}			 &	$\beta$\tablenotemark{\it a}	&$E_p$\tablenotemark{\it a}		&	$E_c$\tablenotemark{\it b}	 & CSTAT/DOF\tablenotemark{\it c}		&	 $\Gamma_{\rm LAT}$\tablenotemark{\it d}				 \\
s			&		&				&				&keV		&MeV	&  &			\\    \hline
52	-	80		&Band &	-0.74	$\pm$	0.02		&	-2.34	$\pm$	0.06  	&	361	$\pm$	12	 &-   &600/348   &	3.49	$\pm$	0.28 	\\
...		&BandCut &	-0.73	$\pm$	0.01		&	-2.25	$\pm$	0.02  	&	355	$\pm$	 6	&224	 $\pm$	 58   &575/347	   &	... 	\\
90	-	160		&Band  &	-1.30	$\pm$	0.03		&	-2.25	$\pm$	0.11  	&	292	 $\pm$	 37	&-   &714/348  &	2.12	$\pm$	0.13 	\\

\hline
\end{tabular}
\tablenotetext{\it a}{The spectral parameters of Band model, $\alpha$ is the photon index below the peak energy of $E_p$, above which the photon index is $\beta$.}
\tablenotetext{\it b}{The high-energy exponential cutoff energy.}
\tablenotetext{\it c}{The Castor Statistic (CSTAT) value and the degree of freedom (DOF).}
\tablenotetext{\it d}{Photon index of the LAT data only.}
\end{table*}

\subsection{KeV/MeV-GeV correlation and the variability of LAT FV component}
\subsubsection{method}
First, the keV/MeV-GeV correlation between the GBM energy flux ($f_{\rm L}$) and
LAT energy flux ($f_{\rm G}$) is tested in a certain time period from $T_1$ to $T_2$
with several time bins, i.e., ${\rm N_{bin}}$. Assuming $f_{\rm L}(T_i)$ and $f_{\rm G}(T_i)$
are the LAT flux and GBM flux at the time of $T_i$, the linear equation can be represent as:
\begin{equation}
f_{\rm L}(T_i) = A + B \times f_{\rm G}(T_i)
\end{equation}
According to Pearson's correlation coefficient $R$ can be represent:
\begin{equation}
R =  \frac{\sum_{1}^{\rm N_{bin}}(f_{\rm G}(T_i)-\bar f_{\rm G})(f_{\rm L}(T_i)-\bar f_{\rm L})}
{\sqrt{\sum_{1}^{\rm N_{bin}}(f_{\rm G}(T_i)-\bar f_{\rm G})^2}  \sqrt{\sum_{1}^{\rm N_{bin}}(f_{\rm L}(T_i)-\bar f_{\rm L})^2}}
\end{equation}
where the $\bar f_{\rm L}$ and $\bar f_{\rm G}$ represent respectively the average fluxes of the LAT band and the GBM band in the chosen time interval. We also calculate the $p$ value of the null hypothesis using the software of {\it Origin}, which can be described as the confidence level of $1-p$ for the keV/MeV-GeV correlation. A strong correlation can be claimed when $R>0.8$ while a moderate correlation can be claimed when $0.5<R<0.8$~\citep{persman-method}. In above two cases, the $p$ value should be smaller than $0.05$, which represents a 95\% confidence level of the correlation. Since the $R$ value in each fit is larger than $0$ ($R=0$, no correlation) in our analysis, thus other cases are defined as a weak correlation.
We first test the correlation in the whole prompt emission duration, which covers
the time period of both GBM and LAT detection (labeled as ``Trace'' hereafter).
Second, we try to determine whether the correlation exists in some sub periods, such as in the FV components.

Second, the ratio $\mathcal{L}$ is calculated between the duration of the full width at half-maximum (FWHM) of the SBPL-fit result and the whole duration of the FV component. The FWHM duration $\omega$ is derived by time spanning a half of the SBPL peak flux in the LAT light curve. The period $T$ (=$T_2-T_1$) is the lower-limit value of the duration of the FV component, which often lasts a longer duration than $T$ as shown in ~Fig~.\ref{fig1} and Fig~.\ref{fig4}.
We regard it as a rapid variability of the LAT FV component if (1) the post-peak decay index $\alpha_d$ being sharper than $-3.0$, since such a sharp decay index cannot be explained by external forward shocks~\citep{2003ApJ...597..455K,2005Sci...309.1833B}, and (2) the ratio $\mathcal{L}$ being smaller than 1.0 significantly, which implies a rapid variability timescale~\citep{2011ApJ...729..114A}.

\subsubsection{Result}

First, we test the correlation between the LAT (0.1-10~GeV) and
GBM (10~keV-10~MeV) light curves during the prompt phase. In the Trace period,
$R$ is $0.90$ with $p <10^{-4}$, which implies a strong correlation
between the GBM and LAT emission. In both the Period 1 and Period 2, the strong
correlations are also found with $R$ of $0.95$ and $0.82$ respectively
, as shown in Tab.~\ref{tab5}.

Second, the correlations of the LAT emission (0.1-10~GeV) and two sub energy bands of GBM emission, i.e., 8-200~keV and 200~keV-1~MeV, are tested.
For the case of 8-200~keV, the LAT-detected emission correlates with that detected in GBM band moderately in the Trace period while in the Period 1 and Period 2, they show a moderate and strong correlation respectively, which can be found in Tab.~\ref{tab5}.
As for 200~keV-1~MeV, strong correlations between the emission in the LAT and GBM energy band
are found in both the Trace period and the Period 1. However, a weak correlation, with $p$ value of $0.17$ is found in the Period 2 due to non-detection of the 200~keV-1~MeV emission in four time bins of this period.

Third, we study the GeV variability of the two LAT FV components during the Period 1 and Period 2, the results of which are given in Tab.~\ref{tab6}. For the first FV component with $\alpha_d$ of $-23.88\pm14.5$ and $\mathcal{L}<0.09$, it exhibits a rapid variability. For the second FV component, we find it is a rapid variability with $\alpha_d$ of $-10.66\pm7.36$ and $\mathcal{L}<0.60$, although with a large uncertainty on the decay power-law index.

The above results on the temporal correlation of keV/MeV-GeV and the temporal variability of the LAT FV components suggest that the prompt high-energy emission in GRB~170214A cannot be produced in the external shock region, but may share the same internal origin as the GBM emission.

\section{Case for other bright LAT GRBs}
\begin{figure*}
    \centering
     \includegraphics[height=5cm]{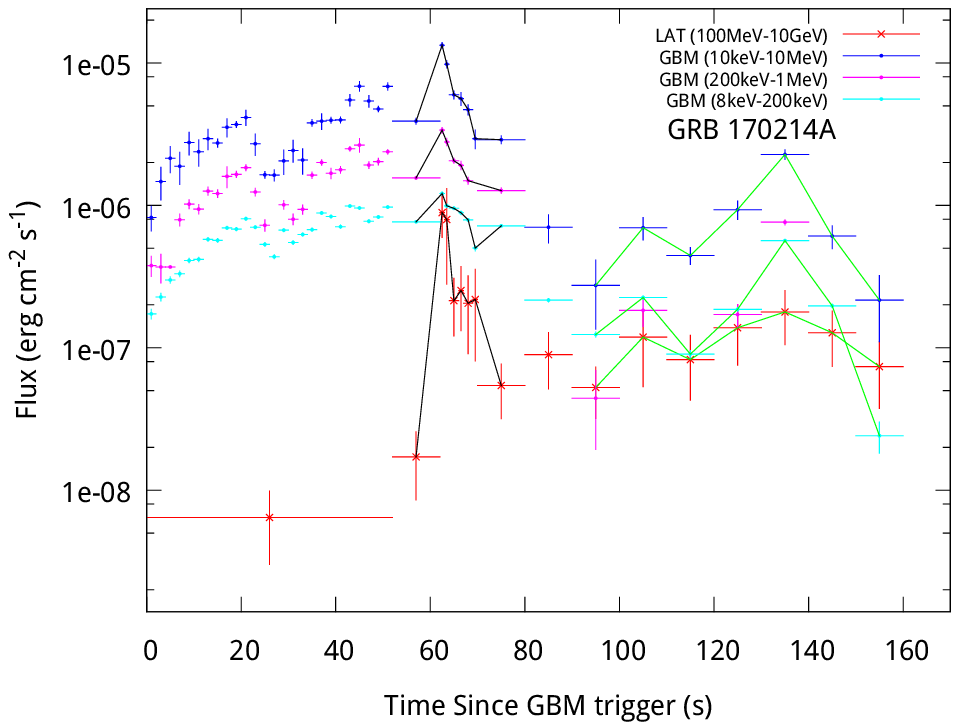}
     \includegraphics[height=5cm]{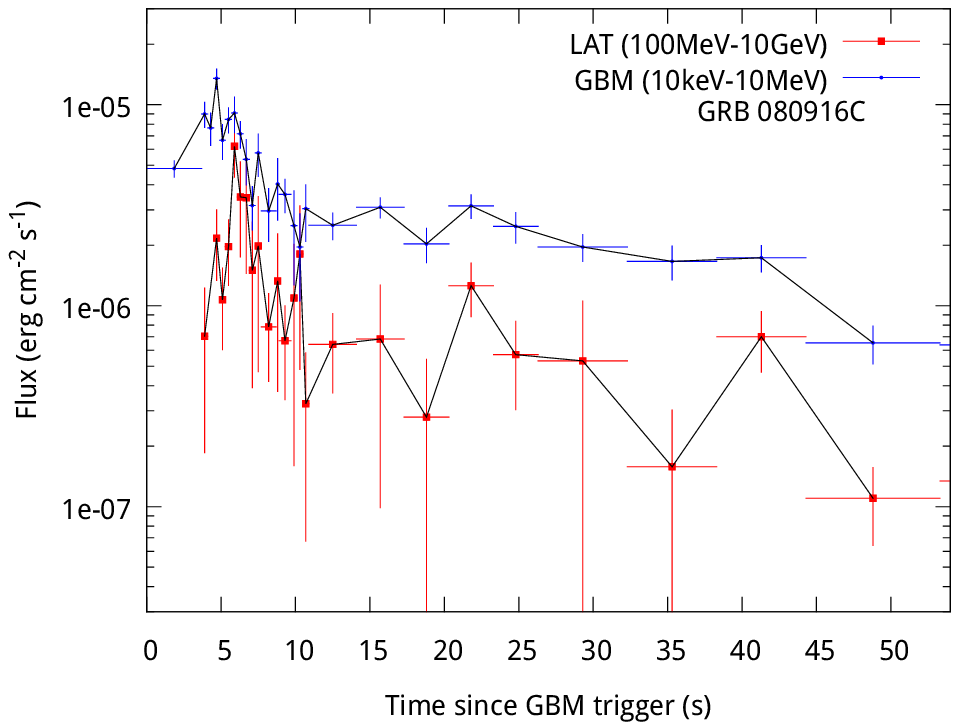}
     \includegraphics[height=5cm]{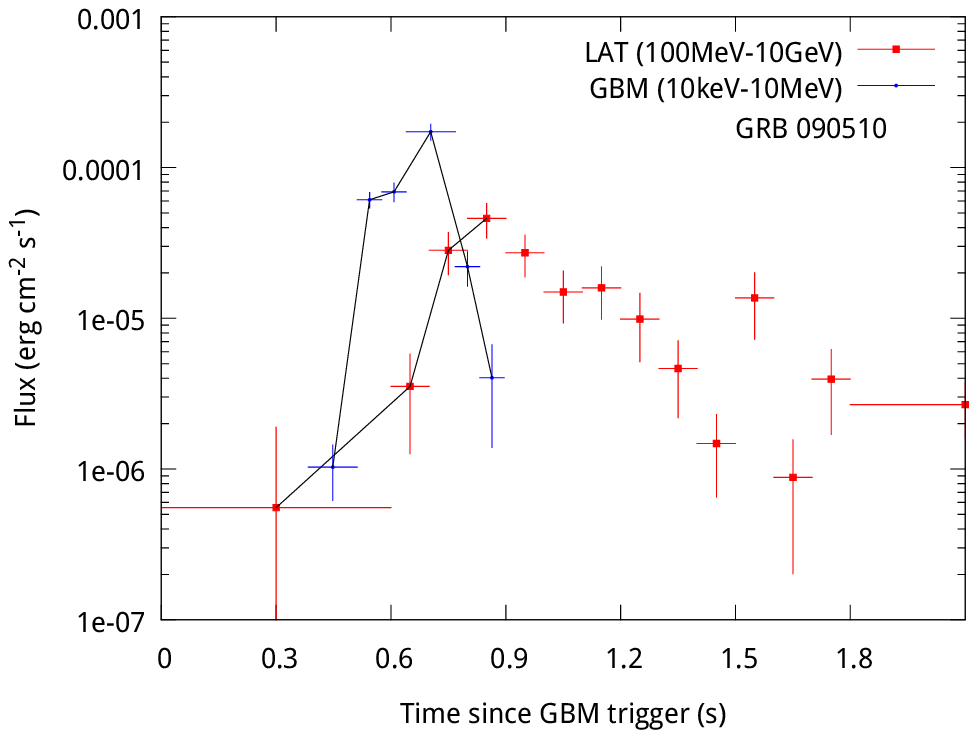}
     \includegraphics[height=5cm]{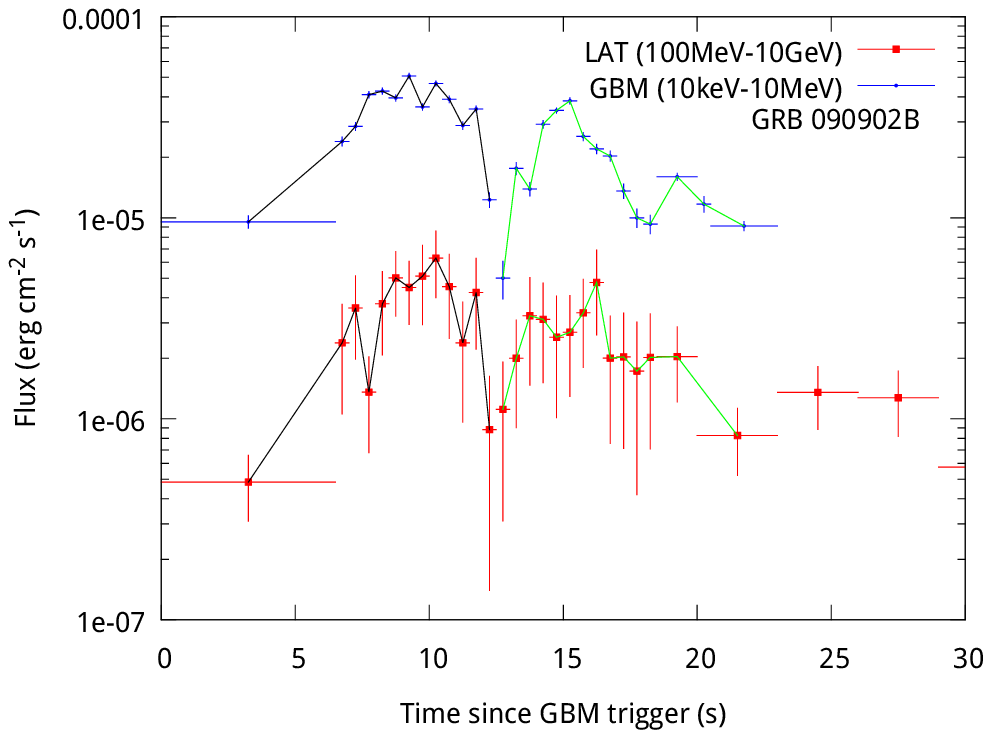}
     \includegraphics[height=5cm]{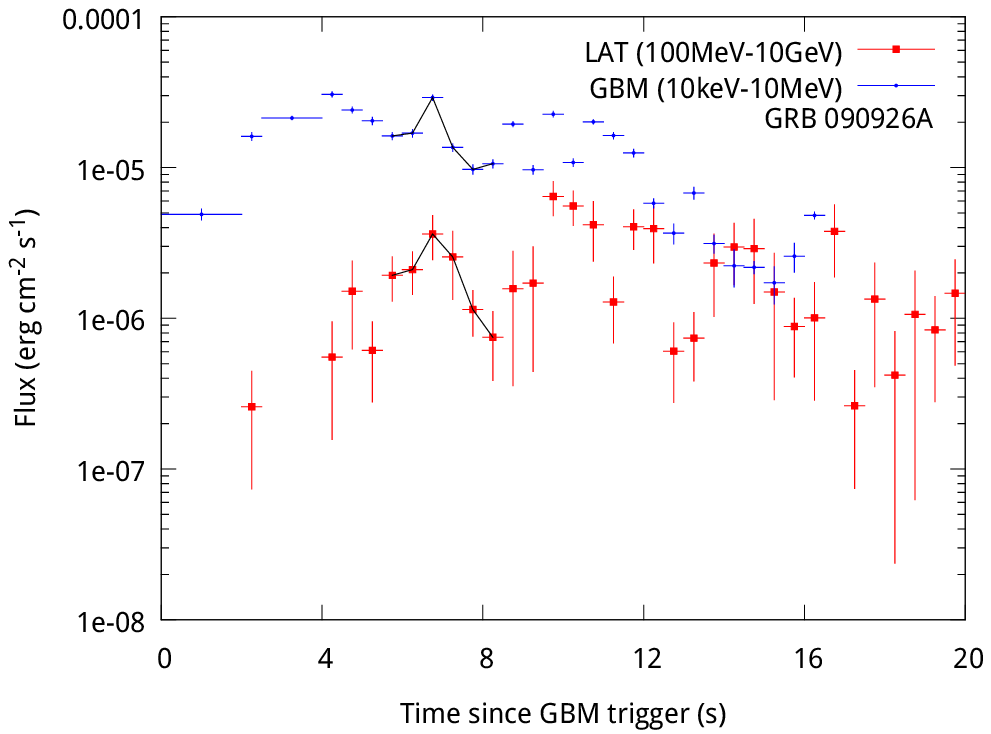}
     \includegraphics[height=5cm]{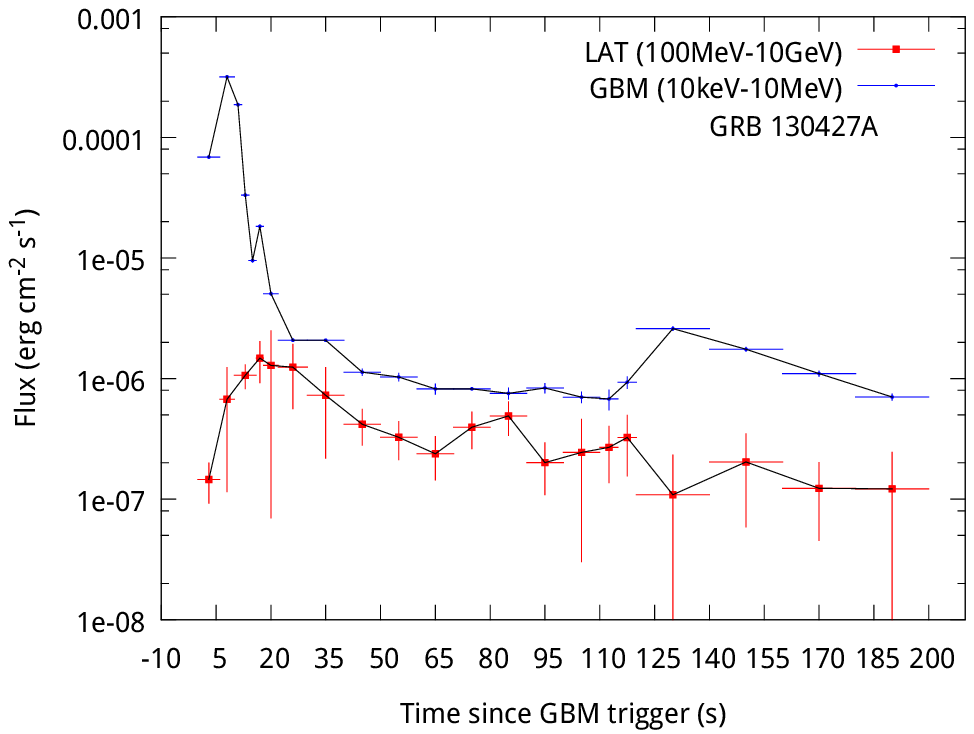}
      \caption{keV/MeV-GeV correlation. (1) GRB 170214A: strong for the Trace period, Period 1 (black line) and Period 2 (green line); (2) GRB 080916C: moderate for the Trace period; (3) GRB 090510: weak for the Trace period; (4) GRB 090902B: moderate for the Trace period, Period 1 (black line) and Period 2 (green line); (5) GRB 090926A: weak for the Trace period, strong for the Period 1 (black line); (6) GRB 130427A: weak for the Trace period.  \label{fig3}}
 \end{figure*}

\begin{figure*}
    \centering
        \includegraphics[height=4.3cm]{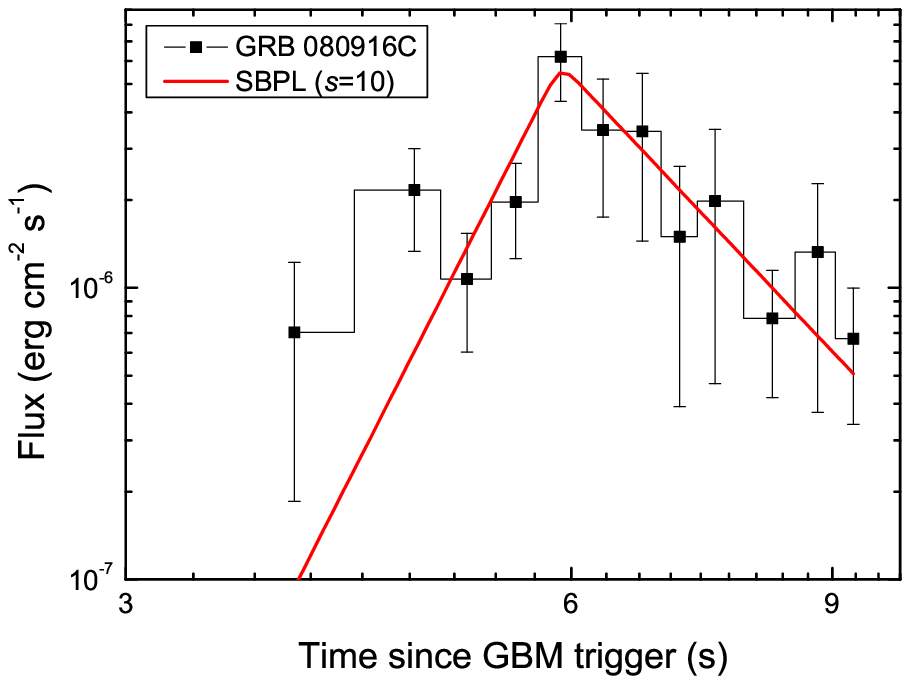}
        \includegraphics[height=4cm]{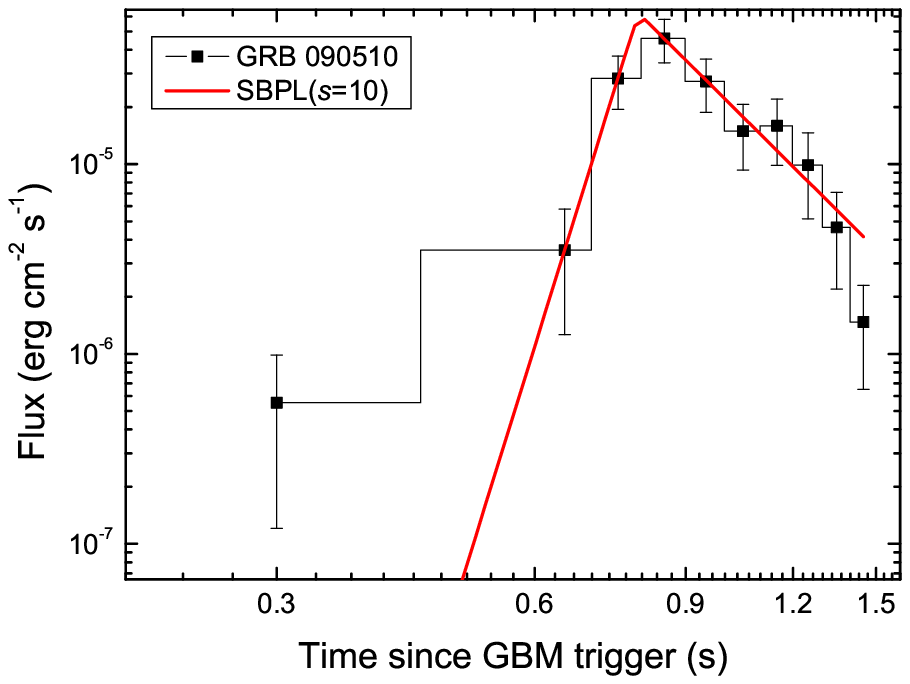}
        \includegraphics[height=4cm]{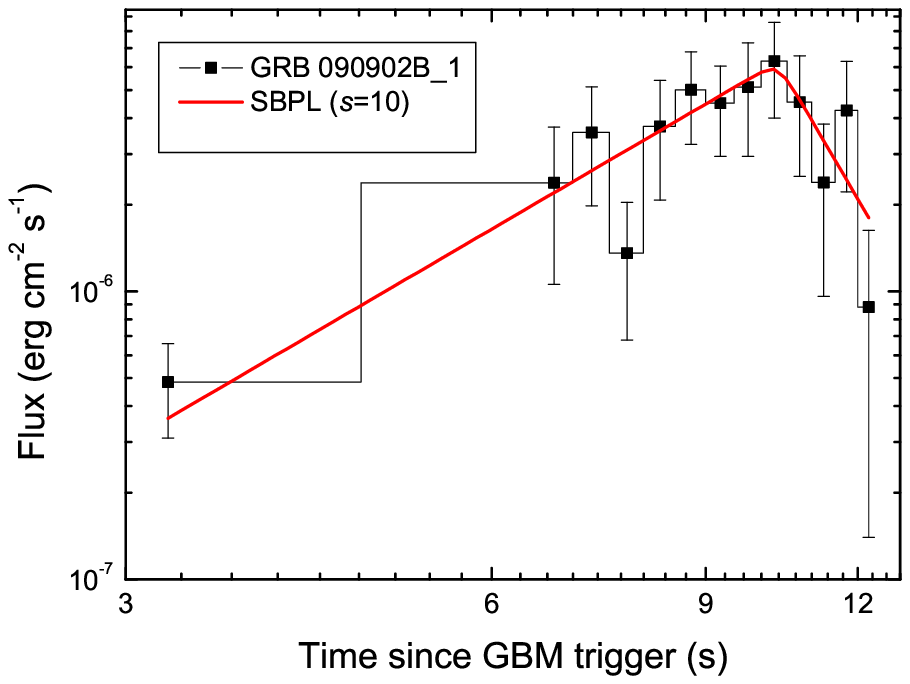}
        \includegraphics[height=4cm]{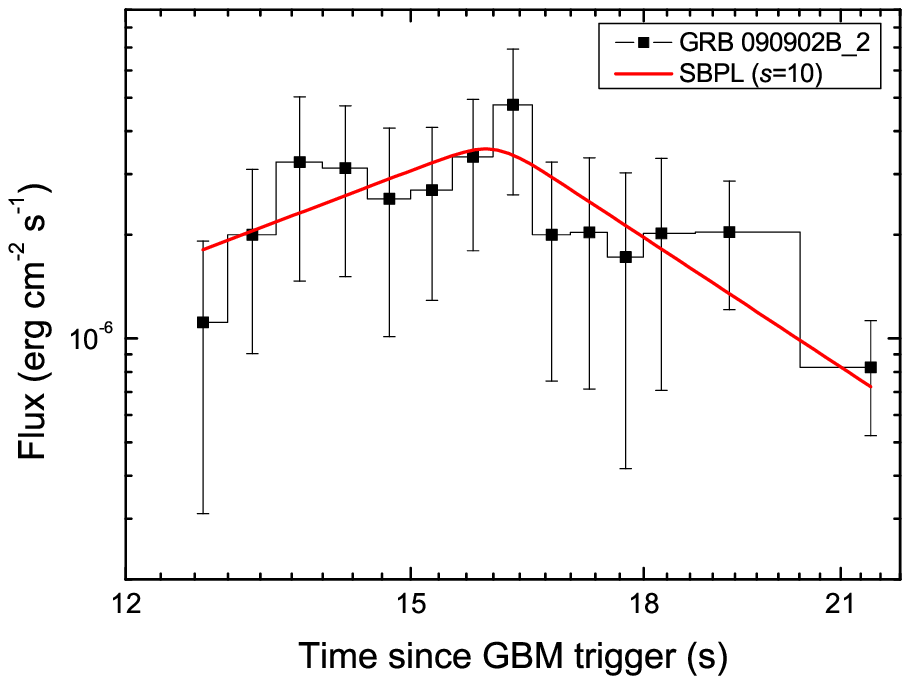}
        \includegraphics[height=4.3cm]{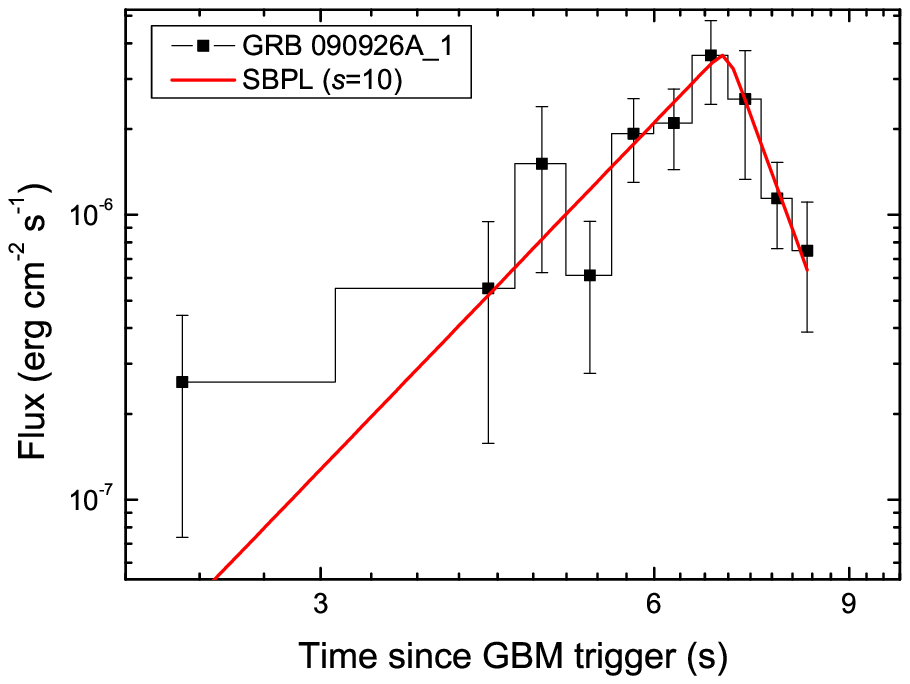}
        \includegraphics[height=4cm]{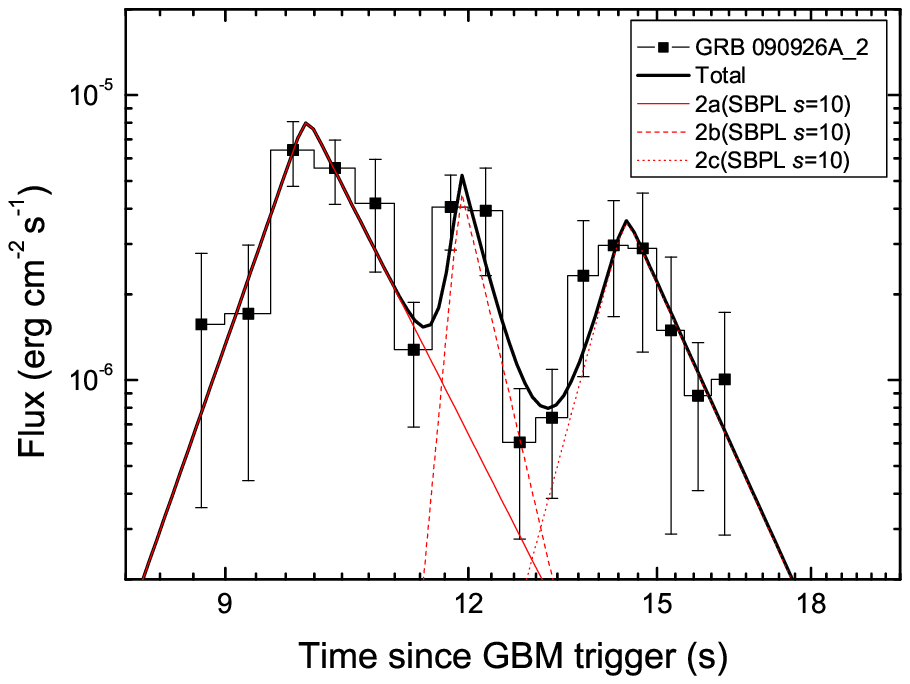}
        \includegraphics[height=4cm]{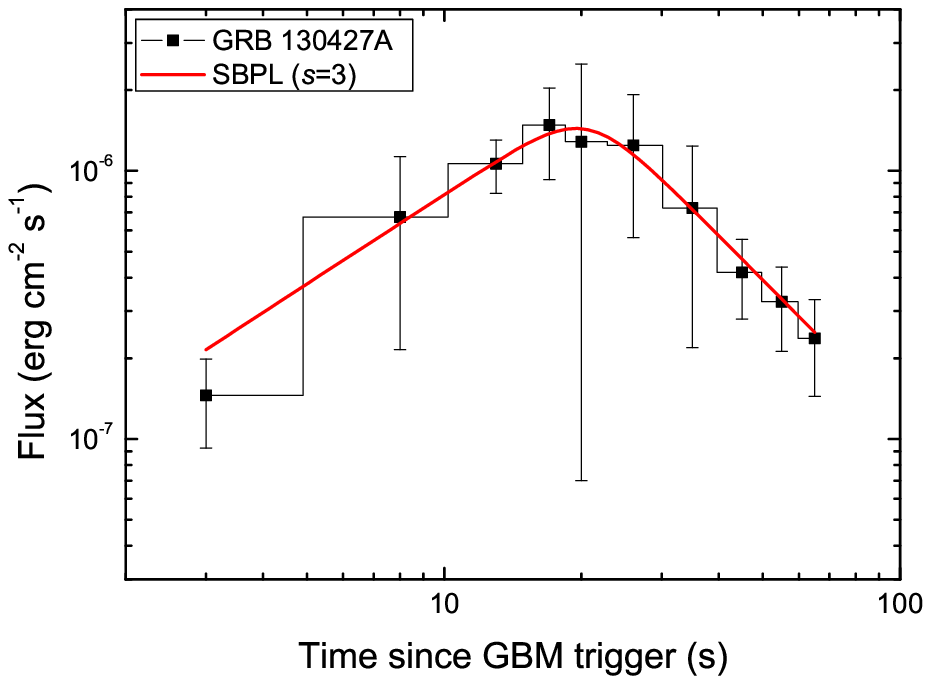}
      \caption{GeV variability analysis for the possible FV components in other five bright LAT GRBs. The filled-square represent the LAT data. The solid line is the best SBPL result of each GRB. For the second FV component of GRB 090926A, three sub structures are found, which are labeled as ``2a'', ``2b'', ``2c'' respectively. The $s$ value is the corresponding smoothness for the SBPL function. \label{fig4}}
\end{figure*}

In this section, we study the keV/MeV-GeV correlation and the variability of the possible LAT FV components for other five bright LAT GRBs, i.e., GRB~080916C, GRB~090510, GRB~090902B, GRB~090926A and GRB~130427A. The results are shown in Tab.~\ref{tab5}, Fig.~\ref{fig3}, Tab.~\ref{tab6} and Fig.\ref{fig4}.

\subsection{GRB 080916C}
In the Trace period, the GBM and LAT light curves show
a moderate correlation with $R$ of $0.59$ and $p$ of $0.0026$.

We search the GeV variability in the first 10 seconds, i.e., between $3.7-9.7$ s. The GeV emission with $\alpha_d$ =$-5.38\pm1.83$ and $\mathcal{L}<0.22$ infers a rapid variability in this period. The resultant peak time of $5.9\pm0.2$ is consistent the prominent peak in the time interval of $3.6-7.7$ s~\citep{2009Sci...323.1688A}.

\subsection{GRB 090510}
GRB 090510 is a short GRB. We find that there is a weak correlation in the Trace period, i.e., $0.3-0.9$ s.

The emission in the LAT band show a rapid rising in the Trace period, after which it exhibits a fast decaying. Thus we extended the time interval to a longer period as a possible FV component, i.e., $0.3-1.5$ s. In this time interval, the LAT light curve shows a rapid variability with $\alpha_d$ =$-4.49\pm0.39$ and $\mathcal{L}<0.17$. The discrepancy between the peak times in the GBM and the LAT light curve is about $0.15$ seconds, which is comparable with the time lag between the GBM and LAT light curves derived in \citet{2010ApJ...716.1178A}, i.e., $0.25\pm0.05$ s.

\subsection{GRB 090902B}
Moderate correlations are found between the GBM and LAT light curves in the Trace period, Period 1 ($0-12.5$ s) and Period 2 ($12.5-23$ s).

The LAT light curve is subdivided into two possible FV components for variability analysis, i.e., $0-12.5$ s and $12.5-23$ s, which is same as the Period 1 and Period 2. For the first FV component, the variability is rapid with $\alpha_d$ =$-7.23\pm3.60$ and $\mathcal{L}<0.30$. The peak time of the first FV components ($10.3\pm0.4$ s) is consistent with that discovered in~\citet{2009ApJ...706L.138A}, which is around $9$ s. For the second FV component, the resultant $\alpha_d$ with a large uncertainty, i.e., $\alpha_d=-3.24\pm1.73$, and $\mathcal{L}<0.53$ can be a rapid variability.

\subsection{GRB 090926A}
A weak correlation is found between the GBM and LAT light curves in the Trace period, although both the light curves show the rapid variabilities. However, we find a strong correlation in the first $8.5$ seconds, with $R$ of $0.89$.

The LAT light curve indeed shows a fast variability. Thus two possible FV components are employed to search the GeV variability, i.e., $2-8.5$ s and $8.5-16.5$ s.
For the first FV component, the variability is rapid with $\alpha_d$ of $-10.63\pm4.25$ and $\mathcal{L}<0.26$.  The peak time (7.0$\pm$0.2 s) of the first FV component locates at the time interval ``b'' ($3.3-9.8$ s)~\citep{2011ApJ...729..114A}.
For the second FV component, we find the LAT light curve could be decomposed into sub structures. Three SBPL components are employed to fit the GeV light curve, the results can be shown in Tab.~\ref{tab6}.  All of them exhibit the rapid variabilities with $\alpha_d$ much sharper than $-3$ and $\mathcal{L}<0.39$.  One of the peak times, i.e., 9.9$\pm$0.2 s, is consistent with the peak time in the time interval ``c'' ($9.8-10.5$ s) in~\citet{2011ApJ...729..114A}.

\subsection{GRB 130427A}
In the Trace period, the correlation between the LAT-detected and GBM-detected emissions is weak. This is consistent with the conclusion drawn by~\citet{2014Sci...343...42A}, i.e., the LAT-detected emission does not appear to be temporally correlated with the GBM emission beyond the initial spike at GBM trigger.

We then perform the analysis in the first $70$ seconds of the LAT-detected emission, a possible FV component, to study the GeV variability. $\alpha_d=1.71\pm0.24$ implies that the variability is not rapid enough to support an internal origin in this period.

\begin{table*}
\centering
\caption{Correlation analysis results of GRB 170214A and other five bright LAT GRBs. \label{tab5}}
\begin{tabular}{ccccccccccc}
\hline
GRB Name &GBM($E_1$	-	$E_2$)	&LAT($E_1$	-	$E_2$)    &$T_1$	-	$T_2$	&	Period	&N${\rm _{bin}}$	& $R$	&	$p$	 & Correlation\tablenotemark{\it a}\\
&keV &GeV   &s			&	&	&		&	&	\\
\hline
170214A	&	10	-	10000    &	0.1	-	10	&	52	-	160	&	Trace	&	16	&	0.90 	&	 $<10^{-4}$	&	 Strong	\\	
...	&	...   &		...		&	52	-	80	&	1	&	8	&	0.95	&	$3.7\times10^{-3}$	&	 Strong	 \\	
...	&	...   &		...		&	90	-	160	&	2	&	7	&	0.81	&	$2.9\times10^{-2}$	&	 Strong	 \\	
...	&	8	-	200	&	...   &	52	-	160	&	Trace	&	16	&	0.68	&	$4.0\times10^{-3}$	 &	 Moderate	\\	
...	&	...   &		...		&	52	-	80	&	1	&	8	&	0.75	&	$3.3\times10^{-2}$	&	 Moderate	 \\	
...	&	...   &		...		&	90	-	160	&	2	&	7	&	0.85	&	$1.5\times10^{-2}$	&	 Strong	 \\	
...	&	200	-	1000	&	...   &	52	-	160	&	Trace	&	11	&	0.81	&	 $2.7\times10^{-3}$	&	 Strong	\\	
...	&	...   &		...		&	52	-	80	&	1	&	7	&	0.96	&	$7.4\times10^{-4}$	&	 Strong	 \\	
...	&	...   &		...		&	90	-	160	&	2	&	4	&	0.83	&	0.17	&	Weak	\\	 \hline
080916C	&	10	-	10000	&	0.1	-	10 &	3.7	-	53.3	&	Trace	&	24	&	0.59	&	 $2.6\times10^{-3}$	&	Moderate	\\	
090510	&	10	-	10000	&		...		&	0.3	-	0.9	&	Trace	&	4	&	0.09	&	0.91	 &	Weak	\\	
090902B	&	10	-	10000	&		...		&	0	-	23	&	Trace	&	27	&	0.73	&	 $<10^{-4}$	&	 Moderate	\\	
...	&		...		&		...		&	0	-	12.5	&	1	&	13	&	0.76	&	 $2.6\times10^{-3}$	&	 Moderate	\\	
...	&		...		&		...		&	12.5	-	23	&	2	&	14	&	0.53	&	 $4.6\times10^{-2}$	&	 Moderate	\\	
090926A	&	10	-	10000	&		...		&	2	-	16.5	&	Trace	&	26	&	0.12	&	 0.56	&	Weak	 \\	
...	&		...		&		...		&	5.5	-	8.5	&	1	&	6	&	0.89	&	 $1.7\times10^{-2}$	&	Strong	 \\	
130427A	&	10	-	10000	&		...		&	0	-	200	&	Trace	&	20	&	0.16	&	0.51	 &	Weak	\\	 \hline
\end{tabular}
\tablenotetext{}{{\bf Note:} (1) $R>0.8$ for strong positive correlation ($p<0.05$); (2) $0.5<R<0.8$ for moderate positive correlation ($p<0.05$); (3) $0<R<0.5$ for weak positive correlation~\citep{persman-method}.}
\end{table*}

\begin{table*}
\centering
\caption{GeV variability of the possible FV components in other five LAT GRBs. \label{tab6}}
\begin{tabular}{cccccccccc}
\hline
GRB\tablenotemark{\it a} & $T_1-T_2$  & $\alpha_r$\tablenotemark{\it b} & $\alpha_d$\tablenotemark{\it b}    & $t_p$\tablenotemark{\it b} & $s$\tablenotemark{\it b}  & $\omega$\tablenotemark{\it c} & $\mathcal{L}$\tablenotemark{\it d} &   Variability\tablenotemark{\it e} \\
    &s  &s  &   &   &   &    \\
\hline
170214A$\_1$	&	52	-	80	&	40.00	(fixed)	&	-23.88	$\pm$	14.5	&	62.3	 $\pm$	 1.1	&	10	&	2.4$\pm 0.5$		&	$<$0.09&	Y	\\
170214A$\_2$	&	90	-	160	&	2.96	$\pm$	1.06	&	-10.66	$\pm$	7.36	&	 140.5	 $\pm$	14.9	&	10	&	41.8$\pm 7.0$		&	$<$0.60 &	Y	\\ \hline
080916C	&	3.7	-	9.7		&	10.00	$\pm$	3.85	&	-5.38	$\pm$	1.83	&	5.9	 $\pm$	 0.2	&	10	&	1.3$\pm 0.2$		&	$<$0.22  & Y\\
090510	&	0.3	-	1.5		&	14.55	$\pm$	4.91	&	-4.49	$\pm$	0.39	&	0.8	 $\pm$	 0.02	&	10	&	0.2$\pm 0.02$		&	$<$0.17	& Y\\
090902B$\_1$	&	0	-	12.5		&	2.47	$\pm$	0.84	&	-7.23	$\pm$	3.60	 &	 10.3	$\pm$	0.4	&	10	&	3.8$\pm 0.5$		&	$<$0.30   & Y\\
090902B$\_2$	&	12.5	-	23		&	5.62	$\pm$	2.56	&	-3.24	$\pm$	1.73	 &	 15.9	$\pm$	0.6	&	10	&	5.6$\pm 0.9$		&	$<$0.53   & Y\\
090926A$\_1$	&	2	-	8.5		&	4.04	$\pm$	0.92	&	-10.63	$\pm$	4.25	&	 7.0	 $\pm$	0.2	&	10	&	1.7$\pm 0.3$		&	$<$0.26	& Y\\
090926A$\_2a$	&	8.5	-	11.5		&	19.47	$\pm$	7.61	&	-13.32	$\pm$	5.07	 &	 9.9	$\pm$	0.2	&	10		&	0.9$\pm 0.07$	 &	$<$0.31	& Y\\
090926A$\_2b$	&	11.5	-	13.5		&	72	(fixed)		&	-29.08	$\pm$	11.53	 &	 11.9	$\pm$	0.2	 &	10	&	0.4$\pm 0.05$		&	$<$0.22	& Y\\
090926A$\_2c$	&	13.5	-	16.5		&	25.17	$\pm$	4.89		&	-14.99	$\pm$	4.78	 &	 14.5	 $\pm$	0.1	&	10		&	1.2$\pm 0.1$	  &	$<$0.39	 & Y\\
130427A	&	0	-	70		&	1.10	$\pm$	0.18	&	-1.71	$\pm$	0.24	&	20.5	 $\pm$	1.5	&	3	&	23.0$\pm 1.8$		&	$<$0.33	& N\\
\hline
\end{tabular}
\tablenotetext{\it a}{The subscript represents the index of the FV component.}
\tablenotetext{\it b}{The parameters of smoothly broken power-law function (SBPL), $\alpha_r$ is the rising index below the peak time of $t_p$, above which the decay index is $\alpha_d$, $s$ is the smoothness of the break.}
\tablenotetext{\it c}{Duration of the full width at half-maximum in the light curve of the FV component.}
\tablenotetext{\it d}{Ratio between $\omega$ and $T (=T_2-T_1)$.}
\tablenotetext{\it e}{A rapid variability when the central value of $\alpha_d$ is smaller than $-3$ and $\mathcal{L}<1.0$.}
\end{table*}

\section{Conclusion}

In this work, we performed the temporal and spectral analysis on GRB 170214A, which shown that the LAT and GBM emission may share the same origin. We thus presented a quantitative analysis of the temporal correlation between the prompt keV/MeV and high-energy ($>100\,$MeV) emission of GRB~170214A. Given the strong correlation found in the periods of the fast variable components and the Trace period, we suggested that the prompt high-energy and keV/MeV emission of GRB 170214A may arise from the same process, say, certain internal dissipation process. Such a temporal correlation is also found in some other LAT GRBs, i.e., GRB 080916C, 090902B and 090926A. The rapid temporal variability  found in LAT emission further supports the internal origin of the high-energy emission in these four GRBs as well as GRB 090510. As our work only deals with the prompt high-energy emission in several bright LAT GRBs,  we need more LAT GRBs with high quality data in future to check whether this is a general case for all the GRBs.

\acknowledgments

We thank the anonymous referee for constructive comments,
which helped us to improve the manuscript. We are grateful for John F. Beacom and
Bei Zhou for a helpful revision of this manuscript.
T.Q.W. is supported by the Natural Science Foundation of China
under grants No. 11547029, 11533004, the Youth Foundation
of Jiangxi Province (No. 20161BAB211007), the Postdoctoral Foundation of Jiangxi Province (No. 2016KY17), the Natural Science Foundation of Jiangxi Provincial Department of Education (No. GJJ150077). W.X.Y. is supported by the 973 program
under grant 2014CB845800, the NSFC under grants 11625312 and
11033002.

\facility{\it{Fermi}}


\begin{thebibliography}{}

\bibitem[Abdo et al.(2009a)]{2009Sci...323.1688A} Abdo, A.~A., Ackermann, M., Arimoto, M., et al.\ 2009a, Science, 323, 1688
\bibitem[Abdo et al.(2009b)]{2009ApJ...706L.138A} Abdo, A.~A., Ackermann, M., Ajello, M., et al.\ 2009b, \apjl, 706, L138
\bibitem[Ackermann et al.(2011)]{2011ApJ...729..114A} Ackermann, M., Ajello, M., Asano, K., et al.\ 2011, \apj, 729, 114
\bibitem[Ackermann et al.(2010)]{2010ApJ...716.1178A} Ackermann, M., Asano, K., Atwood, W.~B., et al.\ 2010, \apj, 716, 1178
\bibitem[Ackermann et al.(2013)]{2013ApJS..209...11A} Ackermann, M., Ajello, M., Asano, K., et al.\ 2013, \apjs, 209, 11
\bibitem[Ackermann et al.(2014)]{2014Sci...343...42A} Ackermann, M., Ajello, M., Asano, K., et al.\ 2014, Science, 343, 42
\bibitem[Band et al.(1993)]{1993ApJ...413..281B} Band, D., Matteson, J.,Ford, L., et al.\ 1993, \apj, 413, 281

\bibitem[Beardmore et al.(2017a)]{2017GCN..20691...1B} Beardmore, A.~P., D'Ai, A., Melandri, A., et al.\ 2017a, GRB Coordinates Network, 20691, 1

\bibitem[Beardmore et al.(2017b)]{2017GCN..20679...1B} Beardmore, A.~P., D'Ai, A., Melandri, A., et al.\ 2017b, GRB Coordinates Network, 20679, 1

\bibitem[Beloborodov et al.(2014)]{2014ApJ...788...36B} Beloborodov, A.~M., Hasco{\"e}t, R., \& Vurm, I.\ 2014, \apj, 788, 36
\bibitem[Burrows et al.(2005)]{2005Sci...309.1833B} Burrows, D.~N., Romano, P., Falcone, A., et al.\ 2005, Science, 309, 1833
\bibitem[Corsi et al.(2010)]{2010ApJ...720.1008C} Corsi, A., Guetta, D., \& Piro, L.\ 2010, \apj, 720, 1008
\bibitem[De Pasquale et al.(2010)]{2010ApJ...709L.146D} De Pasquale, M., Schady, P., Kuin, N.~P.~M., et al.\ 2010, \apjl, 709, L146

\bibitem[Frederiks et al.(2017)]{2017GCN..20678...1F} Frederiks, D., Golenetskii, S., Aptekar, R., et al.\ 2017, GRB Coordinates Network, 20678, 1

\bibitem[Ghirlanda et al.(2010)]{2010A&A...510L...7G} Ghirlanda, G., Ghisellini, G., \& Nava, L.\ 2010, \aap, 510, L7
\bibitem[Gao et al.(2009)]{2009ApJ...706L..33G} Gao, W.-H., Mao, J., Xu, D., \& Fan, Y.-Z.\ 2009, \apjl, 706, L33
\bibitem[Ghisellini et al.(2010)]{2010MNRAS.403..926G} Ghisellini, G., Ghirlanda, G., Nava, L., \& Celotti, A.\ 2010, \mnras, 403, 926
\bibitem[Kobayashi \& Zhang(2003)]{2003ApJ...597..455K} Kobayashi, S., \& Zhang, B.\ 2003, \apj, 597, 455
\bibitem[Kruehler et al.(2017)]{2017GCN..20686...1K} Kruehler, T., Schady, P., Greiner, J., \& Tanvir, N.~R.\ 2017, GRB Coordinates Network, 20686, 1

\bibitem[Kumar \& Barniol Duran(2009)]{2009MNRAS.400L..75K} Kumar, P., \& Barniol Duran, R.\ 2009, \mnras, 400, L75
\bibitem[Kumar \& Barniol Duran(2010)]{2010MNRAS.409..226K} Kumar, P., \& Barniol Duran, R.\ 2010, \mnras, 409, 226
\bibitem[Liang et al.(2008)]{2008ApJ...675..528L} Liang, E.-W., Racusin, J.~L., Zhang, B., Zhang, B.-B., \& Burrows, D.~N.\ 2008, \apj, 675, 528-552
\bibitem[Mailyan \& Meegan(2017)]{2017GCN..20675...1M} Mailyan, B., \& Meegan, C.\ 2017, GRB Coordinates Network, 20675, 1

\bibitem[Malesani et al.(2017)]{2017GCN..20683...1M} Malesani, D., Krogager, J.-K., \& Ranjan, A.\ 2017, GRB Coordinates Network, 20683, 1

\bibitem[Mattox et al.(1996)]{1996ApJ...461..396M} Mattox, J.~R., Bertsch, D.~L., Chiang, J., et al.\ 1996, \apj, 461, 396

\bibitem[Mazaeva et al.(2017)]{2017GCN..20687...1M} Mazaeva, E., Pozanenko, A., Klunko, E., \& Volnova, A.\ 2017, GRB Coordinates Network, 20687, 1

\bibitem[Newton \& Rudestam (1999)]{persman-method} Newton, R. R., Rudestam, K. E. \ 1999, Your Statistical Consultant: Answers to your Data Analysis Questions. Thousand Oaks, CA: Sage Publications

\bibitem[Racusin et al.(2017)]{2017GCN..20676...1R} Racusin, J.~L., Vianello, G., \& Perkins, J.\ 2017, GRB Coordinates Network, 20676, 1

\bibitem[Razzaque(2010)]{2010ApJ...724L.109R} Razzaque, S.\ 2010, \apjl, 724, L109
\bibitem[Sari et al.(1998)]{1998ApJ...497L..17S} Sari, R., Piran, T., \& Narayan, R.\ 1998, \apjl, 497, L17
\bibitem[Schady et al.(2017)]{grb-grond} Schady, P., Greiner, J., Steinmassl, S. \ 2017a, GRB Coordinates Network, 20680, 1
\bibitem[Schady \& Kruehler(2017)]{2017GCN..20684...1S} Schady, P., \& Kruehler, T.\ 2017, GRB Coordinates Network, 20684, 1

\bibitem[Tang et al.(2015)]{2015ApJ...806..194T} Tang, Q.-W., Peng, F.-K., Wang, X.-Y., \& Tam, P.-H.~T.\ 2015, \apj, 806, 194
\bibitem[Troja et al.(2017a)]{grb-ratir} Troja, E., Butler, N., Watson, A., et al. \ 2017a, GRB Coordinates Network, 20681, 1
\bibitem[Troja et al.(2017b)]{grb-ratir2} Troja, E., Butler, N., Watson, A., et al. \ 2017b, GRB Coordinates Network, 20685, 1
\bibitem[Wang et al.(2010)]{2010ApJ...712.1232W} Wang, X.-Y., He, H.-N., Li, Z., Wu, X.-F., \& Dai, Z.-G.\ 2010, \apj, 712, 1232
\bibitem[Zhang et al.(2011)]{2011ApJ...730..141Z} Zhang, B.-B., Zhang, B., Liang, E.-W., et al.\ 2011, \apj, 730, 141
\end{thebibliography}
\end{document}